# Microplasticity and yielding in crystals with heterogeneous dislocation distribution

Xu Zhang[a*], Jian Xiong[a], Haidong Fan[b], Michael Zaiser[a,c]

[a]*Applied Mechanics and Structure Safety Key Laboratory of Sichuan Province, School of Mechanics and Engineering, Southwest Jiaotong University, Chengdu 610031, China*

[b]*Department of Mechanics, Sichuan University, Chengdu 610065, China*

[c]*Institute for Materials Simulation, Department of Materials Science and Engineering, Friedrich-Alexander Universität Erlangen-Nürnberg, Dr.Mack Strasse 77, 90762 Fürth, Germany*

*\*Corresponding author, E-mail: xzhang@swjtu.edu.cn (XZ)*

**Abstract**

In this study, we use discrete dislocation dynamics (DDD) simulation to investigate the effect of heterogeneous dislocation density on the transition between quasi-elastic deformation and plastic flow in face-centered cubic single crystals. By analyzing the stress-strain curves of samples with an initial, axial dislocation density gradient, we arrive at the following conclusions: (i) in the regime of quasi-elastic deformation before the onset of plastic flow, the effective elastic modulus of the simulated samples falls significantly below the value for a dislocation-free crystal. This modulus reduction increases with decreasing dislocation density gradient: crystals with homogeneous dislocation distribution are thus weakest in the quasi-elastic regime; (ii) the transition towards plastic flow occurs first in regions of reduced dislocation density. Therefore, the overall yield stress decreases with increasing dislocation density gradient; (iii) crystals with dislocation density gradient exhibit a more pronounced hardening stage during which stress is re-distributed onto stronger regions with higher dislocation density until the sample flows at a constant flow stress that is approximately independent on dislocation density gradient. We interpret these findings in terms of a continuum dislocation dynamics inspired model of dislocation density evolution that accounts for inversive dislocation motions. The transition between quasi-elastic and plastic deformation is interpreted as a transition from inversive to non-inversive dislocation motion, and the initial differences in elastic modulus are related to a density dependent polarizability of the dislocation system. The subsequent plastic flow behavior is analyzed in terms of a modified version of Mughrabi's composite model.

1. Introduction

Spatially graded microstructures can be observed in many biological materials, such as bones, bamboo, shells, etc., where the microstructure change gradually. These graded structures are the result of natural selection and natural evolution, and almost all of them have some special excellent properties [1]. Therefore, inspired by the gradient biological material, metallic materials with graded microstructures were also fabricated and have shown excellent mechanical properties, fatigue properties, friction and wear behavior [2, 3]. Gradients of microstructural properties may exist on the levels of grain size distribution, twin density, dislocation density, solute or precipitate density, or combinations thereof [4]. There are many studies of the effects of grain size gradients on plasticity [5-7], but there are few studies of the effects of dislocation density gradients despite the fact that such gradients are ubiquitous. Because of the loss of dislocations at surfaces, the existence of a dislocation density gradient in the near-surface region and a concomitant dependency of hardening on distance-to-surface can be demonstrated even in uni-axial deformation of macroscopic samples [8, 9]. Even more pronounced are gradient effects in samples where an intrinsically heterogeneous deformation mode imposes a gradient of strain and hence of strain hardening, see e.g. an investigation of dislocation density gradients of pure nickel under torsion, where deformation and dislocation density gradients were found to be accompanied by a decrease in Vickers hardness from surface to center [10, 11].

The purpose of our study is to understand the effect of dislocation density gradients on the small-strain deformation of a face-centered cubic (fcc) metals by using the discrete dislocation dynamics simulation method to investigate the deformation behavior in the microplastic regime and during yielding. The paper is organized as follows: We first give a concise presentation of the simulation method and present the results obtained by DDD simulation, which we discuss in terms of characteristic differences of dislocation behavior in the quasi-elastic/microplastic and the plastic deformation regime of microcrystals with and without dislocation density gradient. We then move to a more theoretical description, where we formulate a dislocation density based framework to interpret the characteristic differences between dislocation behavior in graded and non-graded samples, and between quasi-elastic and plastic deformation regimes. We conclude with a brief discussion that puts our results into the general context of recent interpretations of the elastic-plastic transition.

2. Simulation Set-up

We use discrete dislocation dynamics (DDD) simulation to describe plastic deformation in terms of dislocation motion and dislocation interactions (see e.g. [12-15]) and apply this method to investigate the effect of dislocation density gradients. Specifically, we use the Parallel Dislocation Simulator (ParaDis) code which uses an adaptive nodal description of the dislocation lines to trace the evolution of the dislocation system [16]. For a detailed description of the code, the reader is referred to the literature [12]. Originally developed by Lawrence Livermore National Laboratory for large-scale hardening simulations [12], ParaDis has become a popular tool for DDD simulation that has been adapted to a wide range of problems including dislocation processes at surfaces [17], size dependent plasticity of small samples [18], and interactions of dislocations with twin boundaries [19].

2.1 Sample geometry

**Table 1**. Material and model parameters used in the DDD simulation and density based model.

| Material parameters | Symbol | Value |
| --- | --- | --- |
| Shear modulus [GPa] | $\mu$ | 54.6 |
| Poisson ratio | $v$ | 0.324 |
| Burgers vector [nm] | $b$ | 0.256 |
| Drag coefficient [Pa · s] | $B$ | $10^{-4}$ |
| Model parameters | Symbol | Value |
| Mean dislocation density [m$^{-2}$] | $\bar{\rho}$ | $7.6 \times 10^{13}$ |
| Mean dislocation source length (nm) | $l_{FR}$ | 400 |
| Fraction of dislocations on active slip systems | $\eta_E$ | 1/3 |
| Taylor constant | $\alpha$ | 0.4 |
| Polarizability for mean dislocation density | $P_0$ | 0.883 |
| Density derivative of polarizability | $P_1$ | 0.0425 |
| Stress re-distribution factor | $\eta_E$ | 0.1 |

In our investigation we consider cuboidal samples of an fcc metal. We use material parameters indicative of Cu as summarized in Table 1. All samples have dimensions $4000b \times 4000b \times 12000b$, which, with a Burgers vector length $b$=0.256 nm, corresponds to sample dimensions of approximately $1\,\mu\text{m} \times 1\,\mu\text{m} \times 3\,\mu\text{m}$. The samples are loaded uni-axially, the load axis is aligned with the [1, -1, 0] axis of a Cartesian coordinate system and the load-perpendicular sample surfaces have [1, 1, 1] and [-1, -1, 2] orientation. This implies that four slip systems on the [1, -1, 1] and [-1, 1, 1] slip planes are loaded symmetrically. These slip systems have a common Schmidt factor of $M = 1/\sqrt{6} \approx 0.408$. The initial configuration and sample geometry are shown in Fig. 1.

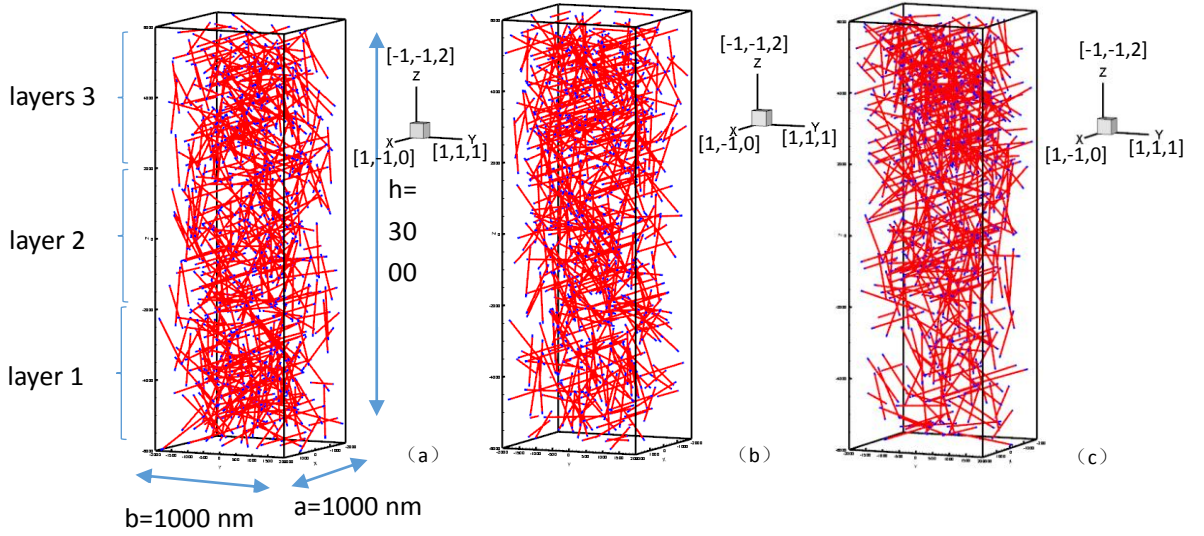

**Fig. 1**. Initial dislocation configurations, blue dots represent F-R dislocation source pinning points, the red lines represent dislocation lines; the coordinate system indicates the crystal orientation of the three samples; (a) sample without dislocation density gradient, (b) low dislocation density gradient, (c) high dislocation density gradient.

2.2 Initial and boundary conditions

Our initial dislocation configuration consists of randomly located Frank-Read Sources, i.e., initially straight segments with pinned endpoints and a source length $l_{FR}$ = 400 nm. We create a total of $n_{FR}$ = 600 sources in the volume $V$ of each simulated sample, which amounts a total dislocation density of

$$\bar{\rho} = \frac{n_{FR} l_{FR}}{V} = 76.5 \times 10^{12} \, \text{m}^{-2} \quad . \tag{1}$$

The sources are equally distributed over the 12 possible slip systems. The orientations of the pinned segments within the respective slip planes are equi-distributed over the unit circle, and the segment centerpoints are located at random positions, under the provision that the segment does not intersect the sample surface.

Sources with pinned endpoints represent artificial configurations and it is necessary to ensure that the artificial pinning points do not control dislocation motion and dislocation multiplication. To avoid artefacts caused by the initial pinning points, it is necessary to ensure that the dislocation source length is significantly larger than the dislocation spacing: In that case, operation of a source is controlled by interactions with other dislocations and by the requirement that the emerging loop can move through the dislocation forest. As discussed in [20], this can be ensured by taking the source length to be at least 2.5-3 times the dislocation spacing, i.e., $\rho l_{\text{FR}}^2 > 10$. This condition is in our simulations always fulfilled. It ensures that the flow stress $\tau_f$ is not mainly controlled by the Orowan stress for source activation, but by the standard Taylor stress representative of forest hardening,

$$\tau_f \propto \alpha \mu b \sqrt{\rho}, \tag{2}$$

where $\alpha \sim 0.2 - 0.5$ is the Taylor factor which depends on the distribution of dislocations over the different slip systems and $\mu$ is the shear modulus of the material.

For the present study about the effect of dislocation density gradients on the initial deformation behavior, the samples are divided along their long axis into three layers of width 4000 $b$ and volume $V/3$. Within each layer, a given number $n_i$ of dislocation sources are located, hence, the dislocation density per layer is

$$\rho_i = \frac{3 n_i l_{FR}}{V}. \tag{3}$$

Dislocation densities and dislocation spacings in each layer are summarized in Table 1 together with the overall dislocation density gradients of the three sample types. Realizations of initial dislocation configurations for the three sample types are illustrated in Fig. 1.

Regarding boundary conditions, we assume that dislocations can freely leave through the sample surfaces. This boundary condition as implemented in standard ParaDis (i.e., without surface corrections to the dislocation stress fields) poses certain problems when compared with simulation schemes that fully account for the elastic boundary value problem [21]. While the stress state for uni-axial deformation is correctly represented on average, the same is not true for the stress fluctuations caused by the intrinsic heterogeneity of dislocation slip. In particular, image forces acting on near-surface dislocations are not accounted for. This implies that, in the near-surface region where image forces prevail over the mutual interactions of dislocations, our simulations do not accurately represent the actual dynamics. This region can be estimated to consist of a layer with a width of about $\bar{\rho}^{-1/2} \approx 0.1$ μm from the specimen surfaces. As a consequence, the results of our simulations, while representing a qualitatively correct picture of the effects caused by surfaces and dislocation density gradients on plastic deformation behavior, cannot be considered fully accurate in quantitative terms. However, for the case of a homogeneous dislocation distribution, the quantitative differences in comparison with comparable simulations that account for the correct surface boundary conditions [22] are minor. For the purposes of gaining a qualitative understanding of the phenomena, which is what we are aiming for, the present simplified approach is therefore fully adequate.

### 3. Simulation results

In the discrete dislocation dynamics, the initial configuration has a significant influence on the simulation results. Due to the different initial configurations, the stress-strain curves and the dislocation density evolution are different in the presence of a dislocation density gradient. Fig. 2 shows the stress-strain curves for the sample without dislocation density gradient and for the two samples containing a heterogeneous dislocation distribution, one with a small and the second with a larger dislocation density gradient as specified in Table 2. From the figure, we can see that the stress-strain curves exhibit two distinct regimes. In the regime of small strains, we observe an approximately linear increase of stress with total strain. However, the behavior in this stage is not simply elastic. This can be seen from the fact that the slope of the stress strain curves (shown in the inset of Fig. 2) falls significantly below the theoretical value for an isotropic material with ν = 0.324 and μ = 54.6 GPa, for which we expect an elastic slope of $E = 2μ(1-ν) = 144.4$ GPa. The actual values of the quasi-elastic slope (henceforth: effective elastic modulus $E_{eff}$), however, are close to 120 GPa. Moreover, the effective elastic modulus is found to depend systematically on the dislocation microstructure. From the initial slope of the stress-strain curve, we find an effective elastic modulus, which for the homogeneous dislocation arrangement amounts to 118.8 GPa, for the sample with weak dislocation density gradient to 120.4

GPa, and for the sample with high dislocation density gradient to 123.5 GPa. We can thus conclude that, in the initial quasi-elastic regime of the stress-strain curve, the presence of a dislocation density gradient leads to a stiffer response of the material.

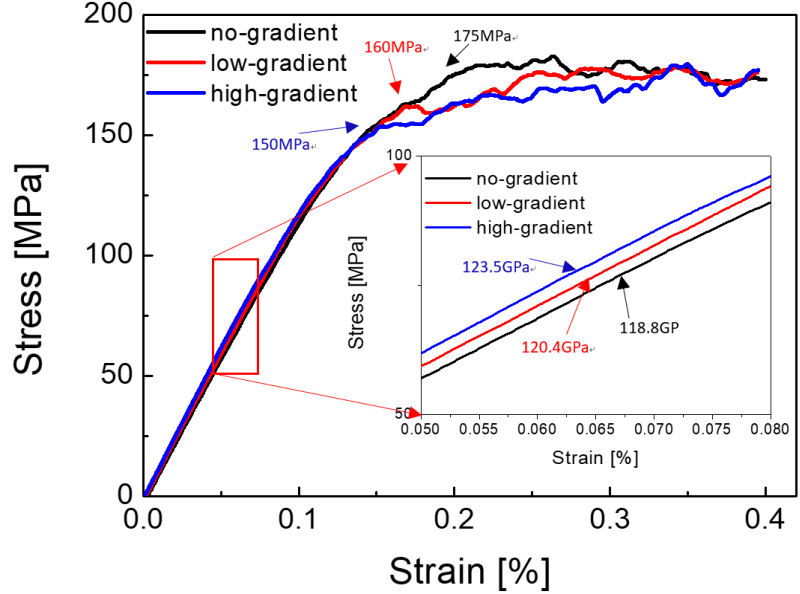

**Fig. 2.** Stress strain curves of samples with different dislocation density gradients. The inset shows part of the quasi-elastic regimes together with the effective elastic moduli of the simulated samples.

**Table 2.** Initial dislocation density distribution for the different sample configurations.

| Cases | No gradient | Low gradient | | | High gradient | | |
|---|---|---|---|---|---|---|---|
| Gradient ($10^{18}$ m$^{-3}$) | 0 | 18.714 | | | 37.428 | | |
| Layers | 1 – 3 | 1 | 2 | 3 | 1 | 2 | 3 |
| Number of sources | 200 | 150 | 200 | 250 | 100 | 200 | 300 |
| Density ($10^{12}$ m$^{-2}$) | 76.5 | 57.4 | 76.5 | 96 | 38 | 76.5 | 114 |
| Dislocation spacing (nm) | 114 | 132 | 114 | 102 | 162 | 114 | 93 |

The opposite behavior is observed during the initial stage of plastic flow. The transition from quasi-elastic to plastic behavior occurs earliest in the material with high dislocation density gradient, later in the sample with small density gradient, and even later in the sample with homogeneous dislocation density distribution.

Accordingly, during the initial stage of plastic flow, the presence of a dislocation density gradient makes the material softer. On the other hand, the initial hardening slope increases with increasing dislocation gradient. As a consequence, the flow stress differences between samples with graded and with homogeneous dislocation microstructure decrease with increasing plastic strain, and above a total strain of about 0.35% (plastic strain of about 0.25%) the flow stress of all samples saturates at a level of about 170 MPa that does not depend on the initial microstructure.

The onset of plastic flow is not associated with any marked increase in the total dislocation density, which, as shown in Fig. 3, changes by less than 10% during the entire simulation. The same is true if we look at the three layers individually (inset in Fig. 3): Even for the largest dislocation density gradient, the dislocation densities in the three layers remain close to their initial values, and there is no systematic dependency of the dislocation density change on the initial dislocation density.

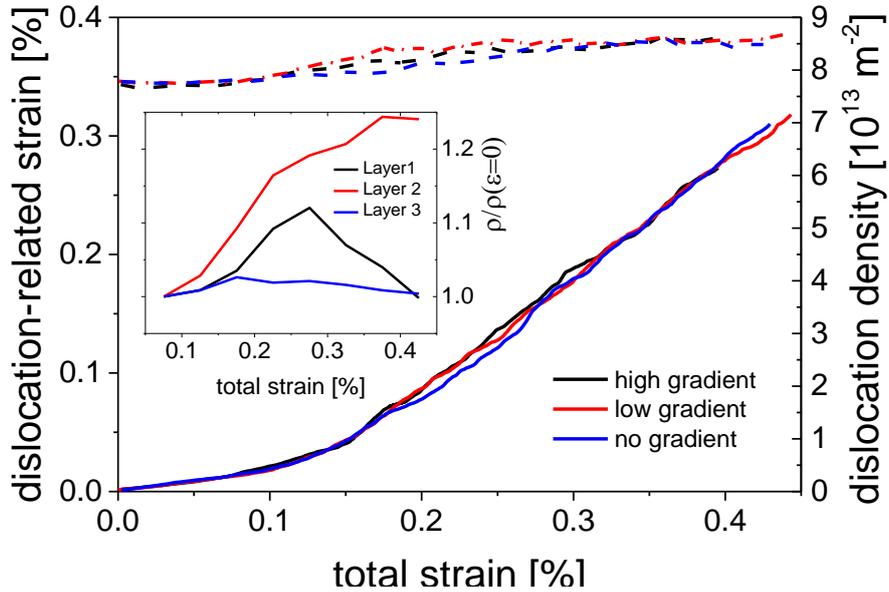

**Fig. 3**. Curves of dislocation-related strain and average dislocation density vs. total strain for the three types of samples; full lines: dislocation-related (inversive plus plastic) strain, dashed lines: total dislocation density; inset: relative change of local dislocation density in the three layers, for sample with high dislocation density gradient.

This finding is remarkable: The simplest explanation for the observed deformation behavior would be that, in line with Eq. (2), deformation starts in the layer where dislocation density and flow stress are lowest. The ensuing local plastic flow then would lead to an increase in dislocation density that, through isotropic hardening, compensates for the initial flow stress differences such that, ultimately, the sample deforms homogeneously and the flow stress reached no longer depends on initial conditions. This explanation is wrong, as the simulations neither indicate a significant increase of dislocation density with plastic strain, nor a reduction of the dislocation density gradient. We will discuss, in Section 4, the reasons for this behavior and propose a better explanation for the observed flow characteristics and dislocation density evolution.

4. Analysis and dislocation density-based modelling
4.1 Quasi-elastic deformation regime

To understand the observed behavior regarding the influence of dislocation density gradients on plastic flow, we first consider the initial regime of quasi-elastic deformation. In this regime, the total strain consists of an elastic contribution, plus a strain that is due to displacement of dislocations out of their configurations of minimum energy. Since this displacement is reverted upon unloading, we cannot call this a plastic strain. On the other hand, because the corresponding motions are associated with energy dissipation, we can also not call it reversible in the thermodynamic sense. We therefore adopt a proposal of Zaiser and Seeger [23] and use the neologism 'inversive' for this type of geometrically reversible, but thermodynamically irreversible dislocation motion. Accordingly, the strain caused by inversive dislocation motions is denoted as the inversive strain $\varepsilon_{\mathrm{inv}}$, and the transition from the initial quasi-elastic to the subsequent plastic deformation regime is understood as a transition from inversive to non-inversive dislocation behavior.

To model the behavior of the dislocation system in the inversive regime, we first consider the paradigmatic case of widely spaced Frank-Read sources that act independently. We use a line tension approximation where we envisage the pinned source segments as elastic lines of line tension $T = Kb^2 \ln(l_{\mathrm{FR}}/b)/(4\pi)$ where $K$ depends on line direction. For Cu, DeWit and Koehler [24] give an approximately sinusoidal dependency of $K$ on the orientation angle $\theta$ with average value $K = 59.3$ GPa. The shape of the bowed out segment is then an elliptical segment (oblate for screw and prolate for edge orientations). We average over all orientations to obtain an average line tension $T = \beta\mu b^2$ with $\beta \approx 0.6$ and a segment shape that is, on average, circular. The radius of curvature of such a segment is $R = \beta\mu b/\tau$, and in the process of bending the segment sweeps the area

$$\Phi(\tau) = \frac{l_{FR}^2}{4u^2}\left[\arcsin(u) - u\sqrt{1-u^2}\right] \quad , \quad u = \frac{l_{FR}}{2R} = \frac{\tau}{\tau_c}. \tag{4}$$

Here $\tau_c = T/(bR_{FR,c}) = 2\beta\mu b/l_{FR}$ is the critical stress for source activation when the radius of curvature reaches the critical value $l_{FR}/2$. For stresses up to 60% of $\tau_c$, Eq. (4) can be well approximated by a linear stress dependency,

$$\Phi(\tau) \approx \frac{2}{3}R^2 u^3 = \frac{1}{12}\frac{l_{FR}^3}{\beta\mu b}\tau. \tag{5}$$

The corresponding strain is inversive as long as the source does not pass its critical configuration. Assuming a system of non-interacting sources of volume density $n_{FR}$, of which a fraction $\eta_A$ is located on active slip systems with non-zero Schmid factor $M$, the inversive strain is in the low-stress regime given by

$$\varepsilon_{inv}(\tau) = M\Phi(\tau)bn_{FR} = \frac{M}{12}\frac{\eta_A n_{FR} l_{FR}^3}{\beta\mu}\tau. \tag{6}$$

We re-write this in terms of the dislocation density, axial stress and Young's modulus as

$$\varepsilon_{inv}(\sigma) = \frac{\eta_A \rho_{FR} l_{FR}^2 (1+\nu)}{36\beta}\frac{\sigma}{E}, \tag{7}$$

where we used that $M^2 = 1/6$. The total strain in the quasi-elastic regime then follows as the sum of the inversive and elastic strains,

$$\varepsilon = \varepsilon_{el} + \varepsilon_{inv} = \frac{\sigma}{E}\left[1 + \frac{\eta\rho l_{FR}^2(1+\nu)}{36\beta}\right], \tag{8}$$

from which the effective elastic modulus is deduced as $E_{eff} = E/[1 + \eta_A \rho l_{FR}^2(1+\nu)/36\beta]$. We thus expect the effective elastic modulus to be reduced due to the presence of inversive dislocation motions.

In quantitative terms, with $\beta = 0.6$ and $\eta = 1/3$ we get an effective modulus $E_{\text{eff}} =$115GPa which is close to the values found in our simulation, though slightly too low. We also note that, if we calculate the average inelastic strain from Eq. (7), the resulting effective elastic modulus is independent of the presence or absence of a dislocation density gradient, in disagreement with the simulation results. Thus, the assumption of independently polarized FR sources needs to be modified.

We can arrive at a better description by assuming that the polarizability of the FR sources is reduced by the influence of the surrounding dislocations by a factor $P$ which describes the reduction, due to stresses of surrounding dislocations, of the effective area between the unstressed source configuration and the saddle-point configuration. We thus replace Eq. (6) by the expression

$$\varepsilon_{\text{inv}}(\sigma) = M\Phi_{\text{eff}}(\tau)bn_{\text{FR}} = \frac{M}{12}\frac{n_{\text{FR}}l_{\text{FR}}^3}{\mu}\tau P(\rho l_{\text{FR}}^2), \tag{9}$$

and accordingly $E_{\text{eff}} = E/[1 + P(\rho l_{\text{FR}}^2)\eta_A \rho l_{\text{FR}}^2(1+\nu)/36\beta]$. (Note that, because of the scale-invariant nature of dislocation interactions [25], the polarizability $P$ cannot depend on the absolute value of $\rho$ but only on the non-dimensional coefficient $\rho l_{\text{FR}}^2$.) For the simulations with homogeneous dislocation arrangement, we obtain from the value of the effective elastic modulus, $E_{\text{eff}} = 118.8$ GPa, the value $P(\bar{\rho}l_{\text{FR}}^2) = P_0 = 0.883$. This leads, for a stress of 100 MPa (total strain 0.085%), to an inversive strain of $\varepsilon_{\text{inv}} = 1.48\times 10^{-4}$ which matches exactly the value deduced from the simulation.

We thus find that interactions between dislocation segments of different sources reduce the polarizability as compared to a system of non-interacting sources. For treating the case of inhomogeneous dislocation arrangements, we expand the polarizability around the reference density $\bar{\rho}$: $P(\rho l_{\text{FR}}^2) \approx P_0 - P_1(\rho-\bar{\rho})l_{\text{FR}}^2$ with p=0.0425. Inserting this relation into the expression for the effective modulus, we find for an inhomogeneous dislocation arrangement:

$$\varepsilon_{\text{inv}}(\sigma) = \frac{\eta_A l_{FR}^2 (1+\nu)}{36\beta} \frac{\sigma}{E} \left( P_0 \bar{\rho} - P_1 l_{FR}^2 (\langle \rho^2 \rangle - \bar{\rho}^2) \right), \tag{10}$$

where the angular brackets denote spatial averages. It is easy to see that, for a heterogeneous dislocation arrangement of a given average density, the mean square density increases with increasing degree of heterogeneity in comparison with a homogeneous system. Accordingly, the inversive strain is reduced and the effective modulus increases. In quantitative terms, we can evaluate the effective elastic modulus of a graded dislocation system as

$$E_{\text{eff}} = E \left[ 1 + \frac{\eta_A \bar{\rho} l_{FR}^2 (1+\nu)}{36\beta} \left( P_0 - P_1 \bar{\rho} l_{FR}^2 \left[ \frac{\langle \rho^2 \rangle}{\bar{\rho}^2} - 1 \right] \right) \right]^{-1}. \tag{11}$$

With $P_1 = 0.0425$ we obtain the effective modulus values in Table 3, which are in good agreement with the simulation data. We may thus conclude that the observation of a stiffer behavior in samples with dislocation density gradient results from the fact that the inversive polarizability of the dislocation system decreases when the ratio between the dislocation spacing and the spacing of pinning points is decreased.

**Table 3.** Effect of dislocation density gradient on effective elastic modulus

|  | Value from simulation | Calculated value from Eq. (11) |
|---|---|---|
| No gradient | 118.8 GPa | 118.8 GPa |
| Low gradient | 120.4 GPa | 120.2 GPa |
| High gradient | 123.5 GPa | 123.9 GPa |

4.2 Plastic deformation regime

The transition towards plastic flow occurs when the externally applied stress 'tilts' the elastic energy landscape to such an extent that dislocations cross barriers in the initial energy landscape, such that upon stress removal they relax into new configurations. In our terminology, this corresponds to a transition from inversive to non-inversive dislocation motions, and from a situation with conserved dislocation density to a situation where both dislocation multiplication and annihilation at the sample surface are present. In order to understand the observed deformation behavior, we thus need to formulate a model describing these processes. To this end, we use the recently formulated approach of continuum dislocation dynamics (CDD) in a version which accounts for dislocation loop generation by FR sources, dislocation line generation by loop expansion, and dislocation losses at the sample surface [26]. We consider the most simple possible CDD version which considers the dislocations on a given slip system in terms of their scalar density $\rho$ and loop density $q$ and note that, because of symmetry, we can treat all four active slip systems in our simulations as symmetry equivalent. Balance equations for $\rho$ and $q$ which account for FR sources have been formulated in [26]. Generalization to include dislocation losses at surfaces is straightforward: Let $A$ denote the cross-sectional area of the specimen volume with an active slip plane. The characteristic rate for a loop to leave the specimen through its surface is given by $v_L = v/\sqrt{A}$ leading to a loop loss rate $\partial_t q^{(-)} = -qv/\sqrt{A}$ and a corresponding dislocation density loss $\partial_t \rho^{(-)} = -\rho v/\sqrt{A}$. In our simulations, $A = 1.22$ and a simple estimate shows that for these values and our dislocation densities, dislocation loss at surfaces prevails over losses due to mutual annihilation. Combining these relations with the generation rates due to loop emission and loop expansion as formulated in [26] leads to the simple evolution equations

$$\frac{\partial \rho}{\partial t} = qv - \rho \frac{v}{\sqrt{A}},$$
$$\frac{\partial q}{\partial t} = \frac{4\pi v_{FR} \rho_{FR}}{l_{FR}^2} - q \frac{v}{\sqrt{A}}. \quad (12)$$

The loop emission rate is controlled by the characteristic velocity $v_{FR}$ of dislocations that are crossing the saddle point configuration of a FR source that is interacting with surrounding dislocations. In a bulk material, as discussed in [27], the velocity at the source is indirectly controlled by the velocity $v$ at which dislocations are convected away from the source as these dislocations exert a back stress, which limits source operation. The same is not true in small samples as studied here, since in these samples the emitted dislocations together with their back stress disappear at the surface sink.

The dislocation velocity $v$ is controlled by collective dislocation interactions, which we express in terms of a dislocation density dependent 'Taylor stress' $\tau_T(\rho) = \alpha\mu b\sqrt{\rho}$. The resulting velocity is given by

$$v = \frac{b}{B}\tau_{\text{eff}} = \frac{b}{B}\left(\tau - \alpha\mu b\sqrt{\rho}\right), \tag{13}$$

whereas the source velocity is additionally influenced by the Orowan stress arising from the need to bend the segment between the source pinning points. Assuming that both effects are additive, we have

$$v_{\text{FR}} = \frac{b}{B}\tau_{\text{eff,FR}} = \frac{b}{B}\left(\tau - \alpha\mu b\sqrt{\rho} - \frac{T}{bl_{\text{FR}}}\right). \tag{14}$$

We can now use this framework to look at the steady state that is reached after some straining when dislocation emission and surface losses mutually balance each other. In this case we can look at steady state solutions of Eq. (12). This gives us the following relations connecting, in the quasi-steady state, the plastic strain rate with the initial FR density, the density of gliding dislocation loops, and the velocities of FR source activation and subsequent glide:

$$\dot{\varepsilon}^{\text{pl}} = M\eta\rho bv = M\eta\rho_{\text{FR}} bv_{\text{FR}} \frac{2\pi A}{l_{\text{FR}}^2}. \tag{15}$$

From this we can infer the FR source velocity in our simulation. We obtain for the parameters in our simulations (homogeneous case) the value $v_{\text{FR}} = 0.048$ m/s. The corresponding effective stress in the saddle-point configuration is negligibly small as compared to the Orowan stress and the Taylor stress. This allows us to set, to a good approximation, $\tau_{\text{eff,FR}} = 0$, $\tau_{\text{eff}} \approx T/(bl_{\text{FR}})$, from which the velocity and density of moving dislocation loops follow as $v = T/(Bl_{\text{FR}}) = 88.5$ m/s and $\rho = 5.4\times10^{11}$m$^{-2}$ which is much less than the dislocation density associated with the source segments.

We can thus conclude, in agreement with the simulation results, that the increase of dislocation density due to the emission of loops is expected to be small, and that the dislocation density remains essentially at its initial level even in the plastic regime. The reason why the dislocation density does not appreciably increase after the onset of plastic flow, and thus that the initial dislocation density gradient does not disappear, lies in the fact that emitted dislocation loops readily disappear at surfaces and therefore source activation does not greatly increase the dislocation density. What then explains the observation that initial flow stress differences due to a dislocation density gradient disappear in the course of deformation?

To answer this question we note that the flow stress is controlled by the stress needed to activate sources, whereas the rate-dependent contribution to the flow stress is small. We can therefore, for a region that is plastically activated, express the local flow stress by the approximate rate-independent relation

$$\tau_\mathrm{f} = \alpha\mu b\sqrt{\rho} + \frac{T}{bl_\mathrm{FR}} \approx \alpha\mu b\sqrt{\rho} + \frac{\beta\mu b}{l_\mathrm{FR}}, \qquad (16)$$

which for a homogeneous dislocation distribution gives us, with the parameters in Table 1, an axial flow stress of $\sigma = \tau_\mathrm{f}/M = 175$ MPa, in good agreement with the simulation data shown in Fig. 2.

To describe the situation of a gradient dependent dislocation distribution with average gradient $\rho' = \partial\rho/\partial x$, we note that in this case the local flow stress is given by

$$\tau_\mathrm{f}(x) = \alpha\mu b\sqrt{\rho(x)} + \frac{T}{bl_\mathrm{FR}} \approx \alpha\mu b\sqrt{\bar{\rho} + \rho'(x-\bar{x})} + \frac{\beta\mu b}{l_\mathrm{FR}}. \qquad (17)$$

where $\bar{x}$ is the midpoint along the specimen axis. To evaluate the corresponding global flow stress, we resort to a variant of Mughrabi's composite model of heterogeneous dislocation density distributions [27]. We assume that the stress is composed of an external stress, which for the present geometry is constant throughout the sample and given by $\sigma_\mathrm{ext} = E(\langle\varepsilon\rangle - \langle\varepsilon^\mathrm{pl}\rangle)$, where the angular brackets denote averages over the sample length, and an internal stress

$$\sigma_\mathrm{int}(x) = -\eta_\mathrm{E} E\left(\varepsilon^\mathrm{pl}(x) - \langle\varepsilon^\mathrm{pl}\rangle\right), \qquad (18)$$

which counter-balances plastic strain heterogeneities. In the original composite model, the Eshelby-like factor $\eta_\mathrm{E}$ is set to unity. This is equivalent to making an iso-strain assumption according to which the total strain ε is constant throughout the sample– an assumption which is warranted in bulk deformation but clearly incorrect in axial deformation of micropillar samples as considered here. We therefore relax this assumption by allowing for incomplete elastic compensation of plastic strain heterogeneities, setting $\eta_E < 1$ to correctly reproduce the observed post-yield stress transients. In the plastic regime, we then require the following inequalities to hold:

$$\begin{aligned} \sigma_{\text{ext}} + \sigma_{\text{int}}(x) &= \frac{\tau_f(x)}{M}, \quad x \in L_p \\ \sigma_{\text{ext}} + \sigma_{\text{int}}(x) &= \frac{\hat{\tau}_f}{M}, \qquad x \in L_e \end{aligned} \quad (19)$$

where $L_p$ denotes the plastically deforming part of the sample, $L_e$ is the part of the sample that undergoes only elastic (or inversive) deformation, and the stress within the plastically inactive region is defined by setting $\sigma_{\text{ext}} + \sigma_{\text{int}}(\varepsilon^{\text{pl}} = 0) = \frac{\hat{\tau}_f}{M}$. These relations allow us to evaluate, the mean (external) stress as

$$\sigma^{\text{ext}} = \frac{1}{MV}\left[\int_{V^{\text{pl}}} \tau_f(x)dx + \hat{\tau}_f V^{\text{el}}\right], \quad (20)$$

the local plastic strain within the active region as $\eta\varepsilon^{\text{pl}}(x) = \varepsilon - \tau_f(x)/(ME)$, the mean plastic strain as $\langle\varepsilon^{\text{pl}}(x)\rangle = (\hat{\tau}_f/M - \sigma^{\text{ext}})/(\eta E)$, and the overall strain as $\langle\varepsilon\rangle = \sigma^{\text{ext}}/E + \langle\varepsilon^{\text{pl}}\rangle$. This leads to the stress-strain curves shown in Fig. 4 (left) which follow the qualitative characteristics seen in the simulations: With increasing dislocation density gradient we observe an earlier transition from elastic to plastic behavior, with plasticity first initiating in the low density regions. Stress re-distribution from regions of low dislocation density (low local flow stress) to regions of high density (high flow stress) then leads to a build-up of internal stresses and kinematic hardening that is the more pronounced the bigger the heterogeneity of the dislocation structure.

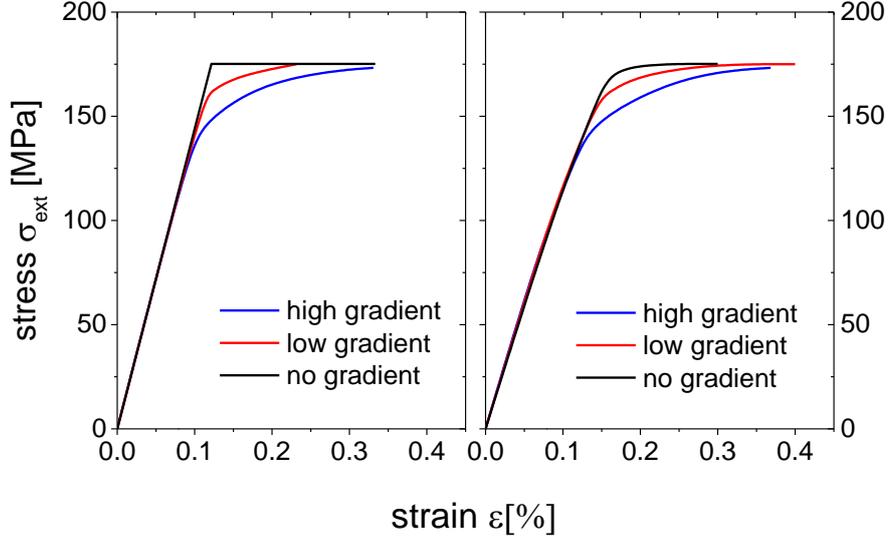

**Fig. 4**. Left: Stress-strain curves computed with modified composite model without accounting for inversive strain contributions, right: stress-strain curves including inversive dislocation strain.

4.3 Elastic-plastic transition

Until now we have disregarded the inversive strain contribution in our discussion of the plastic deformation behavior. In case of a spatially uniform dislocation density this leads to the unrealistic prediction of a quite sharp transition between quasi-elastic and plastic regimes. This shortfall can be rectified by noting that the plastic strain corresponds to the motion of dislocations that have passed the critical saddle point configuration for source activation and then move to the surface, and therefore disregards the inversive strain due to the motion of the same dislocations from their initial equilibrium to the saddle point configuration. For a general inhomogeneous dislocation arrangement, this strain can be evaluated as (see Eqs. (4) and (6))

$$\varepsilon_{\text{inv}}(\tau,\rho) = M \Phi_{\text{eff}}(\tau,\rho) b \frac{\eta_A \rho_{\text{FR}}}{l_{\text{FR}}}, \qquad (21)$$

where now $\tau = M(\sigma_{\text{ext}} + \sigma_{\text{int}})$ and the swept area is evaluated according to

$$\Phi_{\text{eff}}(\tau,\rho) = P(\rho l_{\text{FR}}^2) \frac{l_{\text{FR}}^2}{4u^2}\left[\arcsin(u) - u\sqrt{1-u^2}\right], \qquad (22)$$

where now $u = \tau/\tau_{\text{f}}(\rho)$ if we are in the quasi-elastic regime, and $u=1$ in the plastic regime. Evaluating the ensuing local anelastic strains, integrating them over the specimen length, and adding the result to the elastic and plastic strain contributions gives the stress-strain curves shown in Fig. 4 (right). The curves are now in good agreement with the simulation findings. In particular, we observe even for a homogeneous dislocation arrangement a gradual transition from elastic to plastic behavior, which is associated with a divergence of the inversive polarizability: The stress derivative of the inversive strain, Eqs. (21) and (22), diverges at the critical stress where the system passes its critical saddle point configuration. In the stress-strain curves this implies that the stress approaches the flow stress with a horizontal tangent, but not with a discontinuity of slope.

5. Discussion and Conclusions

We have studied the transition from quasi-elastic behavior to plastic flow in small samples containing either a homogeneous distribution of dislocations or a heterogeneous distribution with a dislocation density gradient. Dislocation motions and dislocation associated strain are already present in the very first stage of deformation, where they are associated with kinematically reversible ('inversive') motions of dislocation segments. In this regime, samples with heterogeneous dislocation distribution exhibit a higher slope of the stress-strain curve. We could interpret this phenomenon in terms of an effective polarizability of the dislocation system that is a decreasing function of dislocation density.

At a critical stress (the flow stress), the polarizability of a homogeneous dislocation system diverges as the system passes over a saddle point and reaches a state of sustained plastic flow. Similar transitions from inversive to irreversibly flowing behavior have been discussed in a wide range of physical contexts, ranging from low-strain deformation of amorphous solids [28] over colloidal suspensions [29], and motion of dislocations under oscillatory stress [30] to vortices in superconductors under oscillating fields [31]. Here we have given for the first time expressions that relate the inversive strain and its diverging susceptibility to basic parameters of the dislocation system in a deforming crystal such as source length and dislocation density.

In the plastically flowing regime we have found that the behavior of the dislocation system in a small sample with open boundaries is governed by the interplay of three factors: (i) activated dislocations (loops emitted from FR sources) rapidly reach the sample surfaces; as a consequence, the dislocation density does not significantly increase and the dislocation microstructure is dominated by the initial, pinned configurations. Accordingly, initially present dislocation density gradients do not decrease during plastic flow. (ii) In heterogeneous dislocation arrangements (density gradients) plastic flow initiates preferentially in the regions of lowest dislocation density, corresponding to lowest local flow stress. Accordingly, the samples with highest dislocation density gradient pass first into the plastic regime. (iii) During heterogeneous plastic deformation, internal stresses re-distribute stress from deformed to less deformed regions. This leads, in samples with dislocation density gradients, to a kinematic hardening stage which is more pronounced if dislocation density and hence flow stress gradients are larger. At the end of this stage, the sample deforms compatibly at a constant stress level which equals the mean flow stress, in line with the prediction of Mughrabi's composite model.

In conclusion, we note that the simulation and model predictions compiled in this paper can be easily checked experimentally. Dislocation density gradients are a ubiquitous feature of the near-surface region of deformed samples [8, 9]. Owing to general scaling relations, such gradients are bound to grow as the overall dislocation density increases in the course of strain hardening. It is therefore possible to prepare surfaces with different dislocation density gradients and then use FIB to produce micropillar samples for controlled testing. It will be an interesting task for future studies to use this technique to gain a clearer understanding of the deformation properties of samples with heterogeneous and graded dislocation structures. This is particularly desirable in view of the technological importance of processes such as shot peening which induce large dislocation density gradients in the near-surface region in order to enhance fatigue and wear resistance.

**Acknowledgements**

This work was supported by National Natural Science Foundation of China (Grant No. 11672251, 11872321, and U1730106), and the Fundamental Research Funds for the Central Universities (Grant No. 2682017QY03). M.Z. also acknowledges support by the Chinese State Administration of Foreign Experts Affairs under Grant No. MS2016XNJT044.